# Effect of the Weakened Heliosphere in Solar Cycle 24 on the Properties of Coronal Mass Ejections

**N. Gopalswamy[1], S. Akiyama[2], S. Yashiro[2], G. Michalek[2,3], H. Xie[2], and P. Mäkelä[2]**

[1]NASA Goddard Space Flight Center, Greenbelt, MD 20771, USA
[2]The Catholic University of America, Washington DC 20064, USA
[3]Astronomical Observatory, Jagiellonian University, Krakow, Poland

nat.gopalswamy@nasa.gov

**Abstract.** Solar cycle (SC) 24 has come to an end by the end of 2019, providing information on two full cycles to understand the manifestations of SC 24 - the smallest cycle in the *Space Age* that has resulted in a weak heliospheric state indicated by the reduced pressure. The backreaction of the heliospheric state is to make the coronal mass ejections (CMEs) appear physically bigger than in SC 23, but their magnetic content has been diluted resulting in a lower geoeffectiveness. The heliospheric magnetic field is also lower in SC 24, leading to the dearth of high-energy solar energetic particle (SEP) events. These space-weather events closely follow fast and wide (FW) CMEs. All but FW CMEs are higher in number in SC 24. The CME rate - sunspot number (SSN) correlation is high in both cycles but the rate increases faster in SC 24. We revisit the study of limb CMEs (those with source regions within 30$^{o}$ from the limb) previously studied over partial cycles. We find that limb CMEs are slower in SC 24 as in the general population but wider. Limb halo CMEs follow the same trend of slower SC-24 CMEs. However, the SC-24 CMEs become halos at a shorter distance from the Sun. Thus, slower CMEs becoming halos sooner is a clear indication of the backreaction of the weaker heliospheric state on CMEs. We can further pin down the heliospheric state as the reason for the altered CME properties because the associated flares have similar distributions in the two cycles – unaffected by the heliospheric state.

## 1. Introduction

Solar Cycle (SC) 24 proved to be the weakest in the Space Age. The reduced solar activity resulted in the weak state of the heliosphere that has severely impacted the observed properties of coronal mass ejections (CMEs) and their space weather consequences. In SC 24, the heliospheric parameters such as the solar wind density, proton temperature, dynamic pressure, and interplanetary magnetic field strength were all diminished compared to the corresponding values in SC 23 [1]. The slow solar wind became quieter with diminished density fluctuations [2]. The total pressure in the heliosphere also decreased, resulting in the anomalous expansion of CMEs in SC 24 [3-6]. One of the clear observational signatures of the anomalous expansion is the slope of the speed-width relationship of limb CMEs in SC 24. The anomalous expansion also resulted in an enhanced abundance of halo CMEs in SC 24: the number did not decline by the same fraction as the sunspot number (SSN) [5]. For a given coronagraph, halo CMEs represent energetic CMEs that expand large enough to surround the occulting disk and hence appear as halo CMEs [7-9]. Although the preferred solar source location of frontside halo CMEs is close to the disk center, a large fraction of halos originated at large central meridian distances (CMDs) in SC24, attributable to the increased CME expansion in the cycle. Many of these results were obtained using CME data from the rise to maximum phase of the two cycles. Now that SC 24 has come to an end by the end of 2019, we check if these results obtained hold for the two complete cycles.

## 2. Overview of CME activity in solar cycles 23 and 24

The Large Angle and Spectrometric Coronagraph (LASCO) [10] on board the Solar and Heliospheric Observatory (SOHO) has generated an unprecedented CME data set that spans SCs 23 and 24. This data set is uniform and uninterrupted except for three months in 1998 and one month in 1999. There have been many studies that compared partial cycles, we now have an opportunity to compare two whole cycles [3-6, 11-12]. In addition, SCs 23 and 24 turned out to be drastically different, providing an opportunity for comparative studies of CME and other activities of the two cycles. For example, the peak SSN, which is used as a measure of the cycle strength, is smaller by 55% (from 180.3 SC 23 to 81.8 in SC 24). The cycle-averaged monthly mean SSN dropped by~40% from 81.1 to 49.1.

One of the early surprises in SC 24 has been a higher CME rate [3, 13-16]. The CME rate is correlated with SSN (see [17] for a review), but the correlation is not perfect because CMEs also originate from non-spot regions [18-19]. As an unfortunate coincidence, LASCO image cadence increased by a factor of ~2 during the rise phase of SC 24 leading to the suspicion that the improved cadence might have resulted in the deviation of CME rate obtained by automatic catalogs from SSN [20-21]. However, the cadence change does not affect the halo CMEs [5]. Higher CME rate is a consequence of the structure and strength of the global magnetic field that changed significantly after the SC-23 polar field reversal [14-15]. All CMEs in the SOHO/LASCO catalog (even those that do not follow SSN) are real and point to the changed global magnetic field structure and the heliospheric state [15].

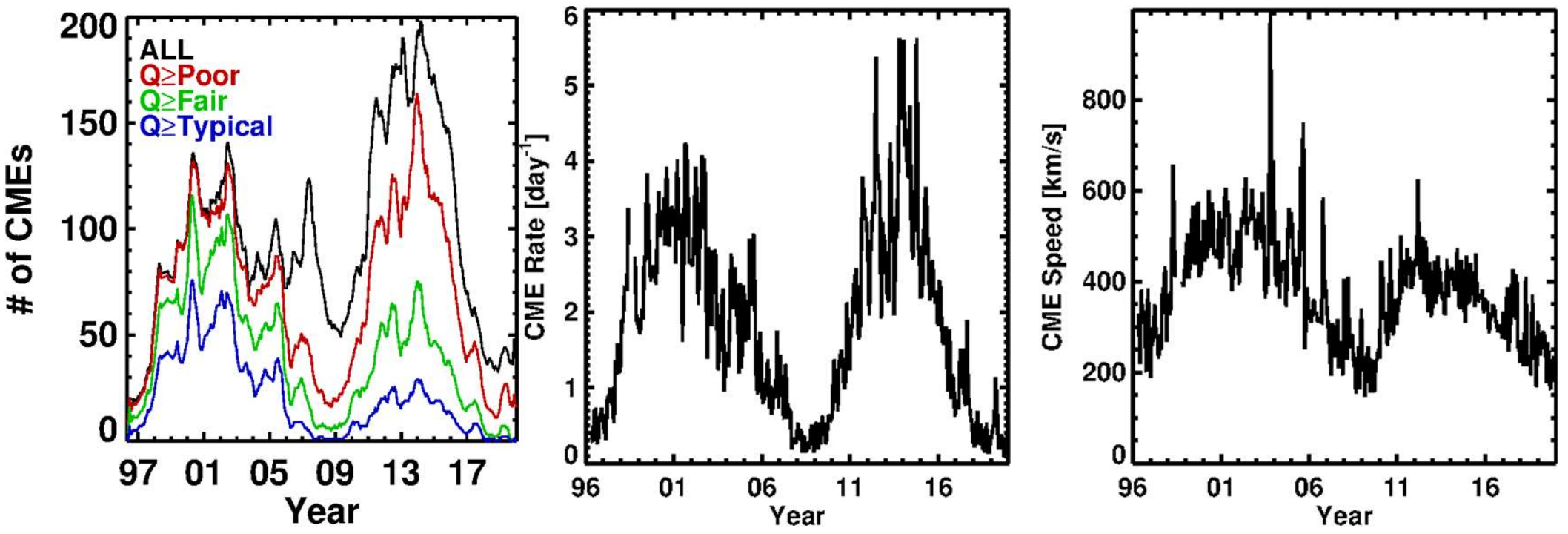

**Figure 1.** (left) Time evolution of CME number per Carrington rotation (CR) counting based on the quality index Q: Ill-Defined (0); Poor (1); Fair (2); Typical (3); Good (4); Excellent (5). "ALL" includes CMEs with all Q's. The Q ≥1 CME number is higher but Q ≥2 number is lower in SC 24. (middle) Daily CME rate (width W ≥30°) averaged over CR periods. (right) Average speed of CMEs in each CR.

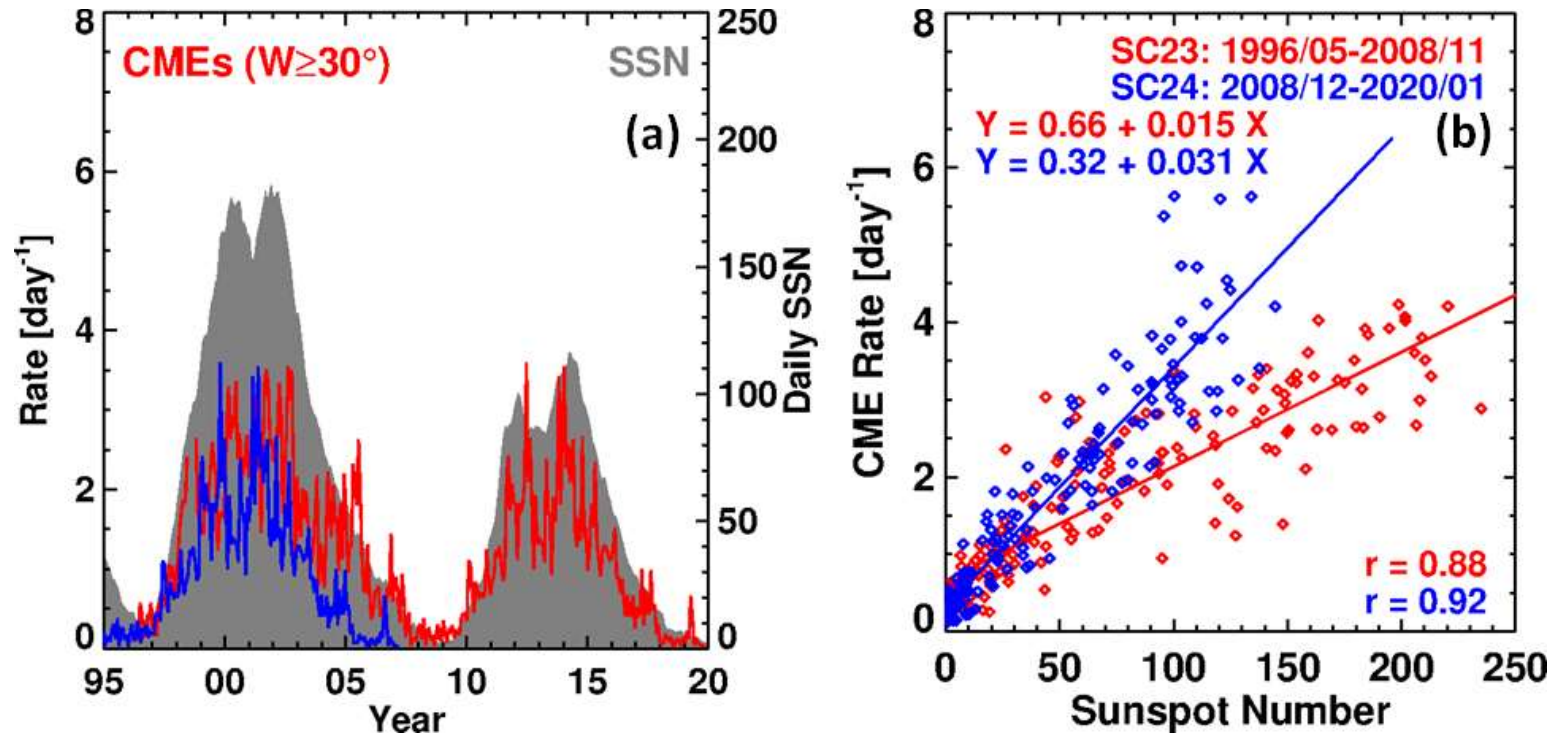

**Figure 2.** (a) The daily occurrence rate of CMEs with W≥30° over the two cycles in red with the SSN in grey (daily SSN smoothed over 13 Carrington rotations; SSN v2 is from Sunspot Index and Long-term Solar Observations – SILSO, http://www.sidc.be/silso/); SC-24 rate is overlaid on SC-23 rate in blue. (b) CME rate vs. SSN (red: SC 23; blue: SC 24) plots showing high correlations, but different slopes (chance-coincidence probability is 0.0 at 95% confidence level).

Figure 2 shows the relation between SSN and CME rate as a function of time. The CME rate clearly tracks the SSN in phase, but not in amplitude. The discordance between CME rate and SSN is pronounced in the maximum phase. SSN and CME rate are well correlated in both cycles with high correlation coefficients (r), but the slope is significantly steeper in SC 24. Steeper slopes are also found in SC 24 when CME rates are compared with other activity indices such as F10.7 [16].

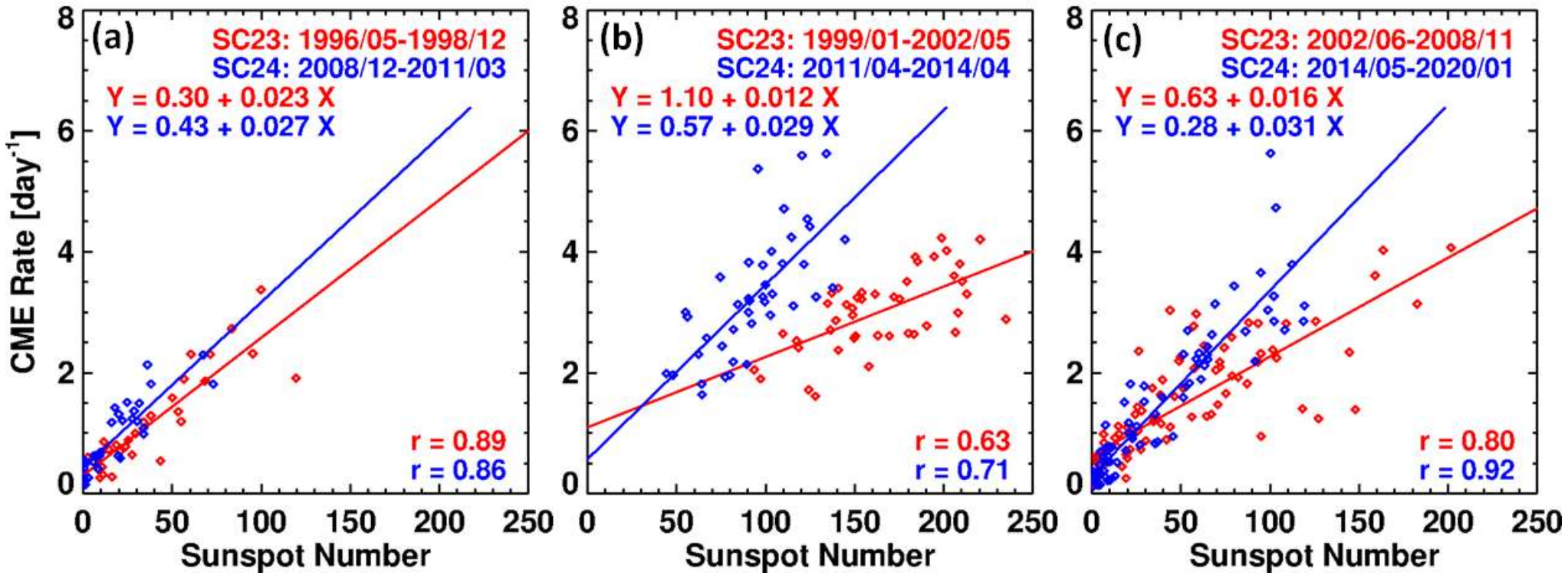

**Figure 3.** SSN (X) vs. CME daily rate (Y) scatter plots in the rise (a), maximum (b), and declining (c) phases in SCs 23 (red) and 24 (blue). CMEs with W ≥30º and Q ≥1 are included in the plots. All correlations are significant. The SC-24 regression lines are steeper in all cases. The slope differences are statistically significant in the maximum and declining phases at 95% confidence level (p=0.00). In the rise phase, p=0.22, so the difference is not significant (too few data points) but the trend is similar to the other phases.

Figure 3 shows a comparison between the corresponding phases in the two cycles. In both cycles, the correlation is weakest during the maximum phase because CMEs from non-spot regions occur in high numbers during this phase [19]. For example, filament eruptions outside the sunspot zones (active region belt) occur in high abundance during the maximum phase (see figure 1 in [24]). In the maximum and declining phases, the difference between the two cycles is statistically significant; in the rise phase, the significance is lower, but has the same trend as in the other two phases.

Table 1 summarizes different sets of CME numbers in the two cycles along with some flare and space weather information. SC 23 started in May 1996 and ended in November 2008, so the cycle length is 151 months. SC 24 started in December 2008 and we take the end to be in December 2019, based on the observations that SC-25 spots have started appearing. This makes the length of SC 24 to be ~133 months. It is known that cycles overlap, so the duration of SC 24 may be a slight underestimate. The CME numbers are obtained from the SOHO/LASCO catalog using the search feature. The number of CMEs in SC 24 is mostly larger except for the fast and wide (FW) CMEs defined as those with speed ≥900 km/s and width ≥60º. The number of full halos and the number of CMEs associated with ≥C3.0 flares near the limb are also smaller in SC 24, but the ratios are larger than that of SSN. Lamy et al. [16] compared CMEs with width >30º that occurred from 1996 until the end of September 2018. They obtained a number ratio (SC-24 count/SC-23 count) as 0.97, which is similar to our value (1.03), given the fact that we considered the full SC 24 until the end of 2019 and CME widths ≥30º. Space weather events (large SEP events, intense geomagnetic storms (Dst ≤-100 nT), ground level enhancement (GLE) in SEPs, and interplanetary (IP) type II bursts observed by the Radio and Plasma Wave experiment (WAVES [25]) onboard Wind are all smaller in number in SC 24. The number of large SEP events and IP type II bursts dropped more than the SSN did. The drop in the number of type II bursts is the same as that of FW CMEs [26]. The drop in the number of large SEP events is expected to be worse because not all FW CME sources are well connected to the SEP observer. The extreme cases are GLE events and large geomagnetic storms that have SC-24 to SC-23 ratios of 0.13 and 0.26, respectively indicating milder space weather in SC 24. These ratios indicate a drop of 87% and 74%, which are significantly larger than the drop in SSN. Both geomagnetic storms and GLE events are due to FW CMEs, which dropped only by 48%. So, in addition to the drop in the number of FW CMEs, other factors such as the heliospheric state also affect the space weather events.

Table 1. Comparison of CMEs, flares, space-weather events, and heliospheric states in the two cycles

| **Property Compared** | **Cycle23[a]** | **Cycle 24[b]** | **Ratio[c]** | |
|---|---|---|---|---|
| | | | **Observed** | **Normalized** |
| Cycle-averaged SSN | 81 | 49 | 0.61 | 1. |
| All CMEs | 13893 (94.5) | 15766 (118.54) | 1.14 (1.25) | 1.88 (2.06) |
| # CMEs W<30° | 5688 (38.69) | 7296 (54.86) | 1.28 (1.42) | 2.11 (2.34) |
| # CMEs W≥30° | 8205 (55.82) | 8470 (63.68) | 1.03 (1.14) | 1.70 (1.88) |
| # CMEs W≥60° | 4163 (28.32) | 4316 (32.45) | 1.04 (1.15) | 1.72 (1.90) |
| # CMEs W = 360° | 396 (2.69) | 324 (2.44) | 0.82 (0.91) | 1.35 (1.50) |
| # CMEs V≥900 km/s & W≥60° | 485 (3.30) | 253 (1.90) | 0.52 (0.58) | 0.86 (0.96) |
| # CMEs ≥C3.0 limb flares | 600 (4.08) | 405 (3.05) | 0.67 (0.75) | 1.11 (1.24) |
| # ≥C1.0 flares | 14730 (97.55) | 8676 (65.23) | 0.59 (0.67) | 0.97 (1.11) |
| Dst ≤ - 100 nT storms | 86 (0.59) | 22 (0.17) | 0.26 (0.29) | 0.43 (0.48) |
| ≥10 pfu SEP events | 102 (0.69) | 46 (0.35) | 0.45 (0.51) | 0.74 (0.84) |
| GLE events | 16 (0.11) | 2 (0.02) | 0.13 (0.18) | 0.22 (0.30) |
| DH Type II bursts | 345 (2.35) | 179 (1.35) | 0.52 (0.57) | 0.86 (0.94) |
| Total Pressure ($10^3$ pPa) | 248 | 185 | 0.75[d] | |
| SW B ($10^2$ nT) | 6.97 | 5.97 | 0.86[d] | |
| SW N ($cm^{-3}$) | 719 | 703 | 0.98[d] | |
| Temperature ($10^5$ K) | 5.5 | 4.2 | 0.76[d] | |
| Alfven Speed (km/s) | 577 | 494 | 0.86[d] | |

[a]From 1996 May 10 to 2008 November 30 (151 months), quantities in parentheses are per month (147 months are used for CMEs in SC 23 to account for the 4-month data gap during the period of SOHO disability); [b]From 2008 December 1 to 2019 December 31 (133 months), quantities in parentheses are per month; [c]Ratio of SC-24 quantity to SC-23 quantity, "observed" denotes the actual ratio, "Normalized" denotes the ratios obtained after dividing the quantities by the cycle-averaged SSN, numbers in parentheses correspond the monthly values. [d]If the solar wind parameters are compared over equal-length periods, the ratios are slightly different: 0.70, 0.82, 0.95, 0.77, and 0.83 from Total pressure to Alfven speed (updated from [4]).

As in previous studies [3,5] that used data from partial cycles, the heliospheric pressure, magnetic field strength, and Alfven speed declined significantly in SC 24. The density drop is small, so the magnetic field essentially decides the change in the Alfven speed. Reduced heliospheric pressure dilutes the magnetic content of CMEs and corotating interaction regions (CIRs) resulting in the reduced geoeffectiveness. Reduction in the heliospheric magnetic field is thought to reduce the efficiency of particle acceleration to high energies, explaining the lack of GLE events. Reduction in Alfven speed, on the other hand, supports shock formation at lower shock speeds and longer lifetime for shocks. However, the drop in the Alfven speed (14%) is about the same as the drop in the speed of CMEs producing IP type II bursts (13%) indicating very little effect on the shock properties (see related discussion on SEP events, [27-29]). However, shocks seem to survive a bit longer in SC 24 based on the higher number of IP type II bursts found at frequencies below 0.5 MHz [26].

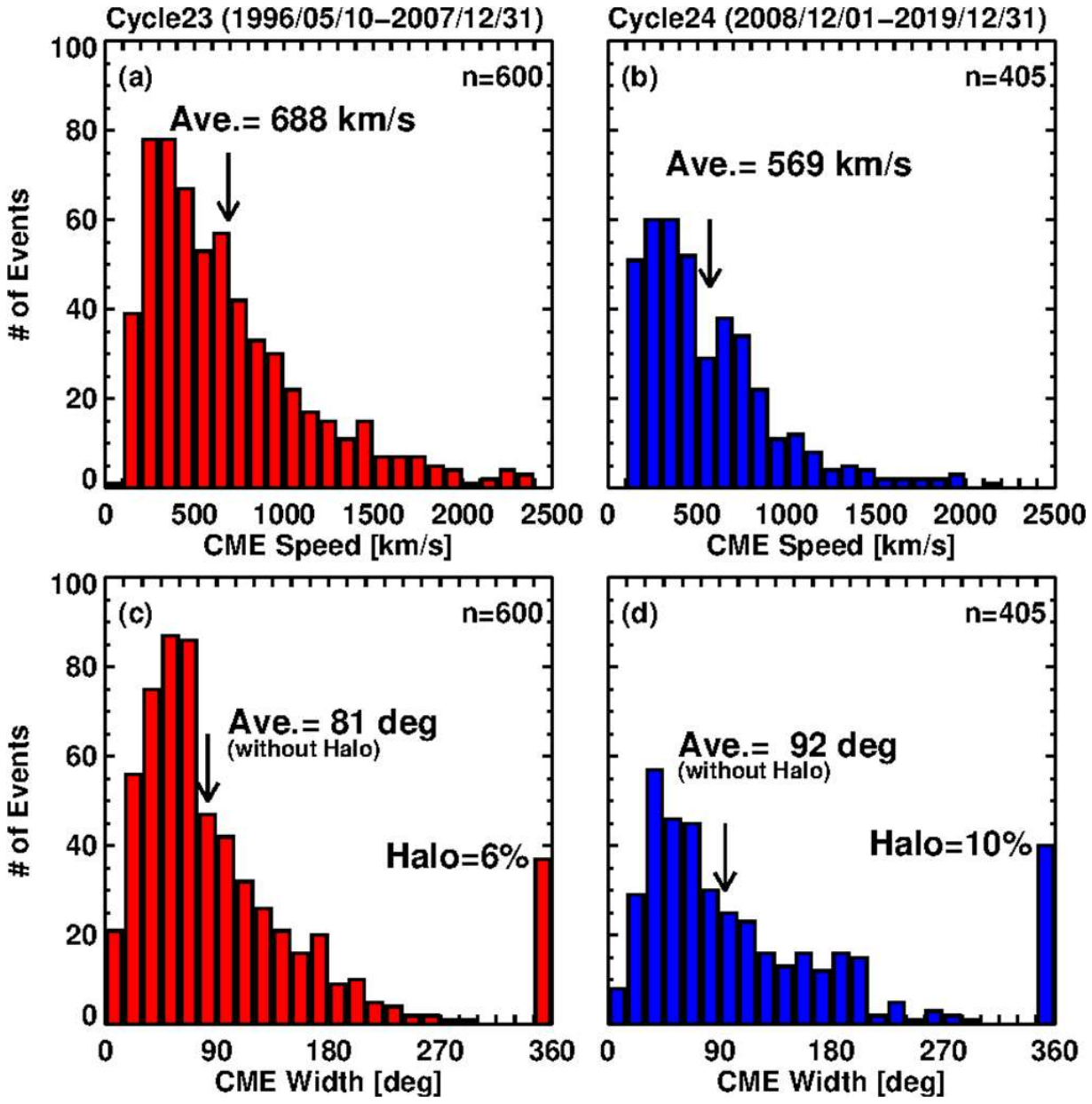

**Figure 4.** Speed (top) and width (bottom) distributions of CMEs associated with limb flares of size ≥C3.0 in SC 23 (a,c) and 24 (b,d). SC-24 CMEs are slower by 13% and wider by 14%. The halo fraction in SC 24 is larger by 67%. A two-sample KS test gives a test statistic D = 0.1220 (p = 0.001), which is higher than the critical value Dc = 0.0875 at 95% confidence level. For the width distributions, D = 0.1054 (p = 0.008), higher than Dc = 0.0875. The non-halo speed distributions are also significantly different with D = 1522 compared to Dc =0.0914. The non-halo widths are different at a lower confidence level (94%).

## 3. Limb CMEs of cycles 23 and 24

Gopalswamy et al. [3] compared CMEs associated with soft X-ray flares of size ≥C3.0 occurring close to the limb (CMD range $60^o$ to $90^o$) during the rise-to-maximum phase of SCs 23 and 24 to find that SC-24 CMEs are wider. The limb CMEs were chosen so that the speed and width measurements can be made with minimal projection effects. Now that SC 24 has come to an end, we have the opportunity to compare limb CMEs in two whole cycles with the sample sizes more than doubled (from 230 to 600 in SC 23 and 148 to 405 in SC 24). The larger samples help us make definite conclusions regarding CME kinematics in the two cycles.

### *3.1 Speed and width distributions*

Figure 4 shows the CME speed and width distributions of the limb CMEs. The average speed in SC 23 (688 ± 38 km/s) is higher than that in SC 24 (569 ± 40 km/s) by 21%, which is much more than the typical error (~10%) in height-time measurements. In the previous study comparing partial cycles, the CME speeds were similar (658 km/s in SC 23 vs. 688 km/s in SC 24) [3]. When only non-halo CMEs are considered, the speeds are also different in the two cycles: 627 ± 33 km/s SC 23 vs. 492 ± 32 km/s SC 24. The whole-cycle comparison finds the SC 24 CMEs to be slower, consistent with the general population shown in Fig. 1. On the other hand, SC-24 CMEs are wider and have a larger fraction of halo

CMEs as was found in [3]. The average width of non-halos in SC 24 is also higher by 14%: 92° ± 6° vs. 81° ± 5° in SC 23.

*3.2 Speed vs. width relationship*

The speed vs. width scatterplots of the ~1000 limb CMEs in Fig. 5 show the different speed and width distributions in the two cycles: the blue and red data points loosely cluster on either side of ~900 km/s, suggesting a larger number of slower and wider CMEs in SC 24. The plot confirms that faster CMEs are wider in general. However, the slope of the SC-24 regression line is steeper (slope is 0.16 vs 0.12 in SC 23), indicating that the SC-24 CMEs are significantly wider for a given speed. The difference in slopes is found to be statistically significant. The difference in slopes found in previous studies [4] continue to hold when larger samples are used. In obtaining the regression lines in Fig. 5, halo CMEs are included. If halos are excluded, the CME numbers decrease to 563 and 365 in SCs 23 and 24, respectively. The slopes of the regression lines decrease to 0.07 (SC 23) and 0.11 (SC 24), still significantly different.

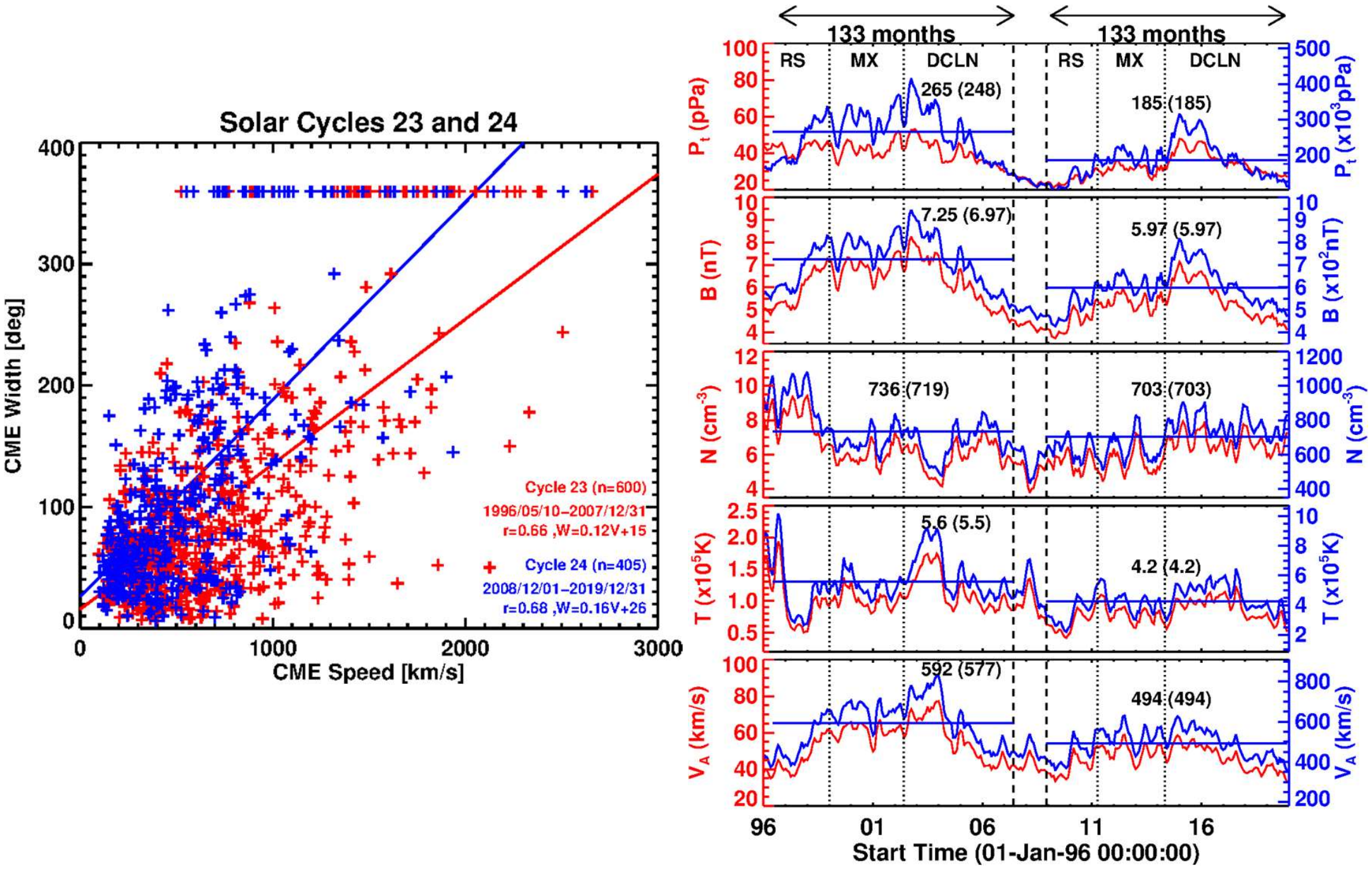

**Figure 5**. (left) Speed (V) vs. width (W) scatter plots of limb CMEs during solar cycles 23 (red) and 24 (blue). While W is correlated with V, SC-24 slope is ~33% steeper. The slope difference is also statistically significant at 95% confidence level ($p=3\times10^{-5}$). The data points at W=360° are halo CMEs. When halos are excluded, the slopes are 0.07 (SC 23) and 0.11 (SC 24) with the difference increasing to 57%. The correlations coefficients (r) are much larger than the Pearson critical coefficients (SC 24: 0.68 vs. 0.163 and SC 23: 0.66 vs.0.138) at 99.9% confidence level. (right) Solar wind parameters at 1 AU (red lines with left-side Y-axis) obtained from the OMNI database from January 1996 through December 2019: total pressure (Pt), magnetic field strength (B), solar wind proton density (N), proton temperature (T), and the Alfvén speed ($V_A$) at 1 AU. Also shown are these parameters extrapolated back to coronal heights (20 Rs), assuming that B, N, and T vary with the heliocentric distance R as $R^{-2}$, $R^{-2}$, and $R^{-0.7}$, respectively (blue lines, right-side Y-axis). The average values over the first 133 months in each cycle are noted. The numbers in parentheses are for the whole cycles. The rise (RS), maximum (MX), and declining (DCLN) phases are delineated; averages in these phases are listed in Table 2.

The plots of solar wind parameters in Fig. 5 confirm that the state of the heliosphere is different in the two cycles, with all the parameters declining in SC 24. The prominent ones are the total pressure determined by the other parameters listed. The total pressure dropped by 25% in SC 24 relative to SC 23 and is responsible for the inflated width of SC-24 CMEs. The Alfven speed declined by 14%, which is the same as the reduction in the magnetic field strength because the density drop is not significant.

Table 2. Solar Wind parameters in the rise, maximum, and declining phases of SCs 23 and 24.

| Property | SC 23 | | | SC 24 | | | Ratio | | |
|---|---|---|---|---|---|---|---|---|---|
| | Rise | Max | Decl. | Rise | Max | Decl. | Rise | Max | Decl. |
| Total Pressure ($10^3$ pPa) | 234 | 301 | 226 | 133 | 192 | 202 | 0.57 | 0.64 | 0.89 |
| B ($10^2$ nT) | 6.74 | 7.85 | 6.61 | 5.07 | 6.12 | 6.26 | 0.75 | 0.78 | 0.95 |
| N ($cm^{-3}$) | 924 | 687 | 652 | 639 | 631 | 679 | 0.69 | 0.92 | 1.04 |
| Temperature ($10^5$ K) | 4.9 | 5.3 | 5.9 | 3.4 | 4.3 | 4.6 | 0.69 | 0.87 | 0.78 |
| Alfven Speed (km/s) | 490 | 656 | 571 | 441 | 536 | 494 | 0.90 | 0.82 | 0.87 |

The average solar wind properties extrapolated to 20 Rs are listed in Table 2. The ratios are similar to that for the whole cycles, but in the declining phase, the ratios are large (less reduction in SC 24). The speed vs. width comparison between corresponding phases in SCs 23 and 24 is shown in Fig. 6. The trend observed for the whole cycles is also observed in the individual phases. The slope difference of the regression lines in the rise and maximum phases are statistically significant; in the declining phase the difference is marginal. These results are consistent with the ratios shown in Table 2.

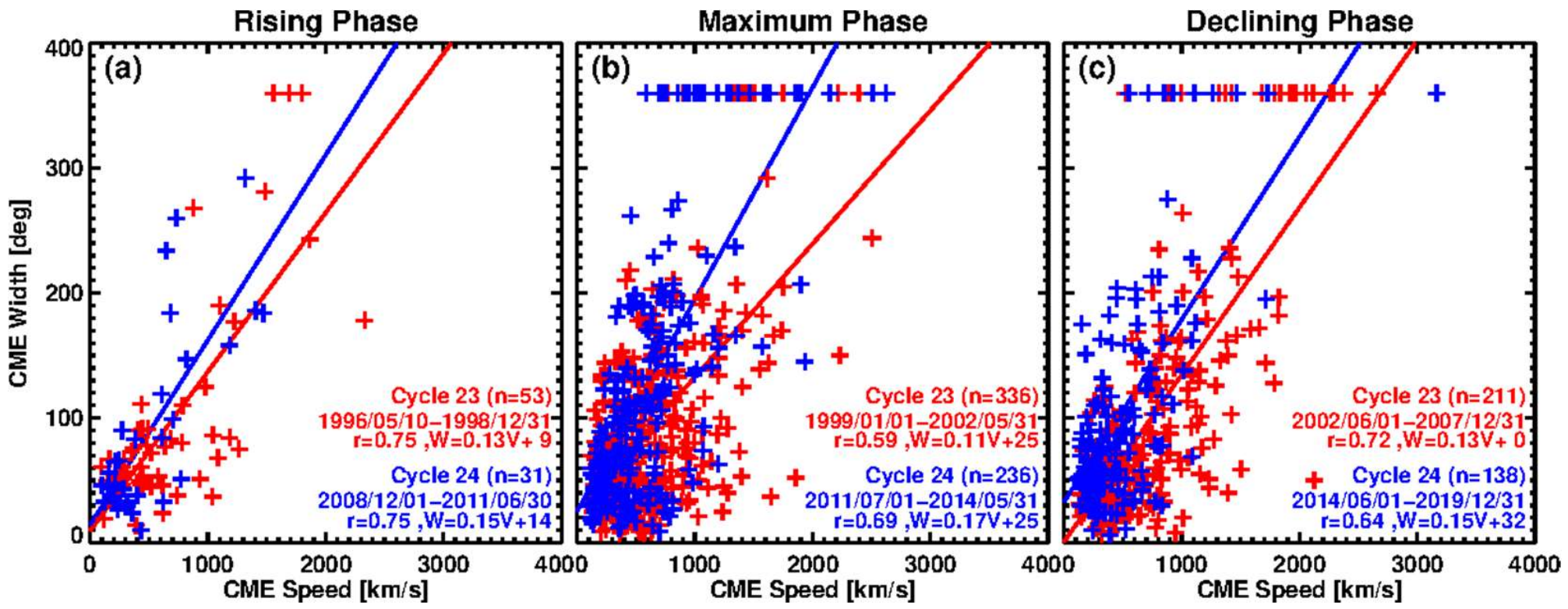

**Figure 6.** Speed (V) vs. width (W) scatter plots of limb CMEs in the rise, maximum, and declining phases in SC 23 (red) and SC 24 (blue). The SC-24 CMEs are wider in all phases. The slope difference between the SC-23 and SC-24 regression lines are statistically significant at 95% confidence level in the rise and maximum phases. In the declining phases, the slope difference has the same trend in the other phases, but the confidence level is low.

*3.3 Flare sizes and active region potential energy*

Our explanation for the wider CMEs and higher halo-CME abundance in SC 24 has been the backreaction of the heliospheric state on CMEs. However, one might ask if the source regions may be responsible for producing faster/wider CMEs (it is well known that there is a correlation between flare size and CME speed). Figure 7 compares the flare-sizes associated with the two sets of limb CMEs. The average flare size in the two populations is nearly the same: M3.7 (SC 23) and M2.7 (SC 24). A two-sample KS test yields a D statistic of 0.0813 ($p = 0.078$), which is smaller than Dc = 0.0875, indicating that there is no statistically significant difference in flare sizes, unlike CME properties. We also computed the magnetic potential energy (PE) of the source active regions that produced the limb flares. The PE is computed from the observed line of sight field strength and area of the active regions at the time of their central meridian passage to avoid projection effects. We counted only unique active regions: regions with multiple flares were counted only once. The distribution of PE in Fig. 7 shows that the average PE in SC 24 ($1.83\times10^{33}$ erg) is smaller than that in SC 23 ($5.09\times10^{33}$ erg) by ~64%. The smaller active region PE is another consequence of the weak SC 24. Typically, it takes several eruptions to exhaust the free energy available in active regions and the resulting flares seem to be similar in the two SCs. Therefore, we rule out the source properties as the reason for the CME width difference.

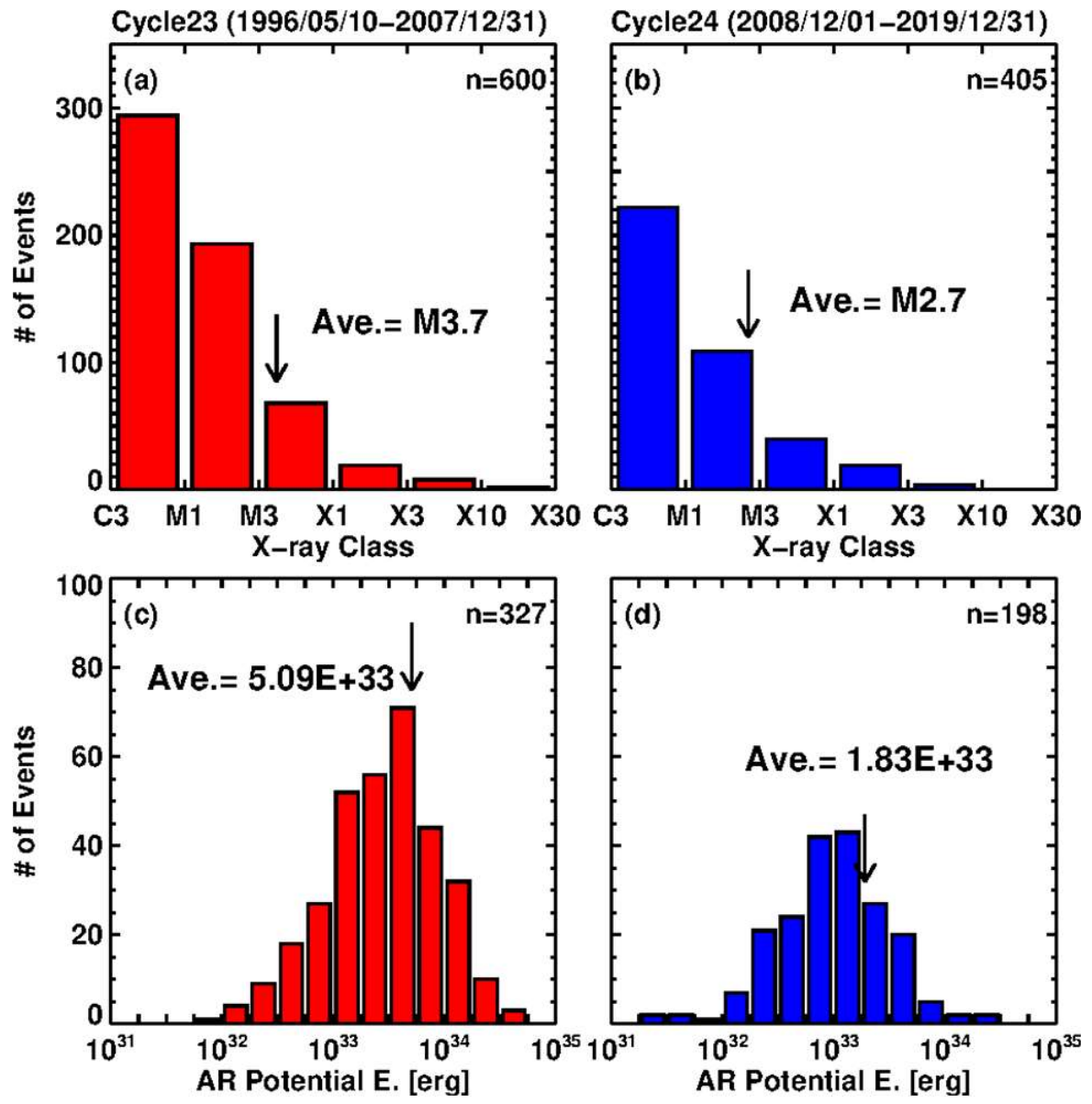

**Figure 7.** Distributions of GOES soft X-ray flare size (top panels) and active region (AR) potential energy in SC 23 (a,c) and SC 24 (b,d). The averages of the distributions are shown on the plots. The potential energy was computed when the flare-producing source region was near the central meridian.

*3.4. Limb halos: a special population*

Limb halos are a subset of limb CMEs in which an eruption on one limb results in an observable disturbance above the opposite limb [30]. The disturbance is likely to be a shock because the flux ropes do not expand >180°. We know this because limb halos occasionally produce interplanetary CME signatures at Earth and geomagnetic storms due to the shock sheath [31-32]. Our previous study on full halo CMEs [5] was concerned mainly with the disparity of high number of halos in SC 24 (irrespective of the source location on the disk). Gopalswamy et al. [30] made use of limb halos (a subset of limb CMEs used in Fig. 5) to show that halo CMEs are formed sooner in SC 24 with lower speeds as a consequence of the weak heliospheric state. To show this quantitatively, they introduced a new parameter known as halo height (the CME leading-edge height when it becomes halo for the first time). The CME nose height is unknown for halo CMEs originating close to the disk center (blocked by the occulter). For limb halos, the projection effects are minimal, so the halo height is accurate. Here we have expanded the list of limb halos reported in [30] by adding 32 behind the limb events whose sources are within 30° from the limb (see Table 3). In SC 23, we determined the eruption locations by extrapolating the location of the underlying source region assuming a solar rotation rate of 13°.2 per day. In SC 24, we used images from the Extreme Ultra Violet Imager [34] on board the Solar Terrestrial Relations Observatory (STEREO) to identify the eruption location. We also used radio and particle data from STEREO to check SEP and type II burst association. Figure 8 shows the 2002 September 6 halo originating from S08W120 in the field of view (FOV) of the Extreme-ultraviolet Imaging Telescope (EIT [33]) and LASCO FOV. The flare was completely occulted. The halo height is 6.18 Rs. In SC 24, there are two halo heights: one obtained with regular cadence and the other (in parentheses in Table 3) with reduced cadence to match the SC-23 cadence.

Table 3. Halo CMEs originating within 30° from the limb in solar cycles 23 and 24

| #[a] | CME Date [UT] | Time [UT] | Location | Flare Size[b] | Halo Height [Rs] | $V_{Sky}$ [km/s] | $V_{Sp}$ [km/s] | SEP ?[c] | Type II [MHz][d] |
|---|---|---|---|---|---|---|---|---|---|
| | | | | **Cycle 23** | | | | | |
| 1 | 1997/11/06 | 12:10:41 | S18W63 | X9.4 | 8.97 | 1556 | 1604 | Y | 14-0.1 |
| 2 | 1998/04/23 | 05:55:22 | S17E90 | X1.4 | 9.62 | 1691 | 1691 | HiB | 14-0.2 |
| 3 | 1998/11/24 | 02:30:05 | S28W90 | X1.0 | 13.40 | 1798 | 1798 | M | 1-0.4 |
| 4 | 1999/07/25 | 13:31:21 | N38W81 | M2.4 | 17.67 | 1389 | 1392 | N | 0.2? |
| 5 | 2000/04/04 | 16:32:37 | N17W60 | C9.7 | 13.57 | 1188 | 1372 | Y | 14-0.2 |
| 6 | 2000/05/05 | 15:50:05 | S17W87 | M1.5 | 10.77 | 1594 | 1594 | M | 14-2.5 |
| 7 | 2000/10/16 | 07:27:21 | N03W90 | M2.5 | 9.78 | 1336 | 1336 | Y | 14-1 |
| 8 | 2000/10/24 | 08:26:05 | S23E70 | C2.3 | 13.24 | 800 | 820 | N | N |
| 9 | 2000/10/25 | 08:26:05 | N09W63 | C4.0 | 7.98 | 770 | 813 | Y | 10-0.3 |
| 10 | 2001/04/01 | 11:26:06 | S22E90 | M5.5 | 13.72 | 1475 | 1475 | HiB | DG |
| 11 | 2001/08/19 | 06:06:05 | N30W75 | B8.4 | 11.18 | 556 | 563 | HiB | 1-0.4 |
| 12 | 2001/10/01 | 05:30:05 | S24W81 | M9.1 | 10.19 | 1405 | 1409 | Y | 1-0.15 |
| 13 | 2001/11/22 | 20:30:33 | S25W67 | M3.8 | 13.55 | 1443 | 1472 | Y | 8-1 |
| 14 | 2001/12/14 | 09:06:06 | N07E86 | M3.5 | 14.22 | 1506 | 1507 | N | 0.7-0.3 |
| 15 | 2001/12/28 | 20:30:05 | S24E90 | X3.4 | 12.63 | 2216 | 2216 | Y | 14-0.35 |
| 16 | 2002/01/04 | 09:30:05 | N38E87 | C3.7 | 12.10 | 896 | 907 | HiB | Nm |
| 17 | 2002/01/14 | 05:35:07 | S28W90 | M4.4 | 11.05 | 1492 | 1492 | Y | 14-0.35 |
| 18 | 2002/02/20 | 06:30:05 | N12W72 | M5.1 | 10.80 | 952 | 965 | Y | 14-10? |
| 19 | 2002/03/10 | 23:06:55 | S22E90 | M2.3 | 12.31 | 1429 | 1429 | N | 14-8 |
| 20 | 2002/03/22 | 11:06:05 | S10W90 | M1.6 | 15.79 | 1750 | 1750 | Y | 14-0.5 |
| 21 | 2002/04/21 | 01:27:20 | S14W84 | X1.5 | 20.25 | 2393 | 2396 | Y | 10-0.06 |
| 22 | 2002/07/19 | 16:30:05 | S15E90 | C2.9 | 11.20 | 2047 | 2047 | HiB | 5-1 |
| 23 | 2002/07/20 | 22:06:09 | S13E90 | X3.3 | 13.74 | 1941 | 1941 | M | 10-2 |
| 24 | 2002/07/23 | 00:42:05 | S13E72 | X4.8 | 25.22 | 2285 | 2318 | HiB | 11-0.4 |
| 25 | 2002/08/22 | 02:06:06 | S07W62 | M5.4 | 8.42 | 998 | 1034 | Y | 14-3.5 |
| 26 | 2002/08/24 | 01:27:19 | S02W81 | X3.1 | 12.09 | 1913 | 1920 | Y | 5-0.4 |
| 27 | 2002/12/08 | 23:54:05 | S18E70 | C2.5 | 13.76 | 1339 | 1361 | N | DG |
| 28 | 2003/05/31 | 02:30:19 | S07W65 | M9.3 | 19.65[c] | 1835 | 1888 | Y | 3-0.15 |
| 29 | 2003/06/15 | 23:54:05 | S07E80 | X1.3 | 14.18 | 2053 | 2062 | N | 14-0.4 |
| 30 | 2003/11/04 | 19:54:05 | S19W83 | X28. | 11.53 | 2657 | 2662 | Y | 10-0.2 |
| 31 | 2003/11/11 | 13:54:05 | S03W61 | M1.6 | DG | 1315 | 1367 | HiB | 1-0.5? |
| 32 | 2004/07/29 | 12:06:05 | N00W90 | C2.1 | 18.00 | 1180 | 1180 | M | 1-0.05 |
| 33 | 2005/01/20 | 06:54:05 | N14W61 | X7.1 | DG | ---- | ---- | Y | 14-0.03 |
| 34 | 2005/06/03 | 12:32:10 | N15E90 | M1.0 | 14.48 | 1679 | 1679 | N | 10-0.27 |
| 35 | 2005/07/13 | 14:30:05 | N11W90 | M5.0 | 15.14 | 1423 | 1423 | M | 14-1 |
| 36 | 2005/07/14 | 10:54:05 | N11W90 | X1.2 | 17.31 | 2115 | 2115 | Y | 3-0.8 |
| 37 | 2005/07/27 | 04:54:05 | N11E90 | M3.7 | 14.08 | 1787 | 1787 | Y | 1-0.45 |
| 38 | 2005/07/30 | 06:50:28 | N12E60 | X1.3 | 9.73 | 1968 | 2043 | Y | 9-0.08 |
| 39 | 2005/08/22 | 17:30:05 | S13W65 | M5.6 | 10.77 | 2378 | 2445 | Y | 12-0.04 |
| 40 | 2005/08/23 | 14:54:05 | S12W70 | M2.7 | 18.45 | 1929 | 1929 | HiB | 13-0.2 |
| 41 | 2005/09/05 | 09:48:05 | S07E81 | C2.7 | 12.50 | 2326 | 2334 | N | 1.5-0.06 |
| 42 | 2005/09/09 | 19:48:05 | S12E67 | X6.2 | 8.37 | 2257 | 2311 | HiB | 14-0.05 |
| 43 | 2006/12/06 | 20:12:05 | S05E64 | X6.5 | DG | ---- | ---- | Y | 14-0.03 |
| 44 | 2007/01/25 | 06:54:04 | S08E90 | C6.3 | 14.97 | 1367 | 1367 | N | 14-0.09 |
| 45 | 1996/12/02 | 15:35:05 | S05W90 | C2.7 | 8.19 | 538 | 538 | N | N |
| 46 | 2001/04/18 | 02:30:05 | S21W116 | C2.2 | 17.78 | 2465 | 2767 | Y | 14-0.1 |
| 47 | 2001/06/15 | 15:56:27 | S07W118 | C1.9 | 13.23 | 1701 | 1945 | Y | 14-3.5 |
| 48 | 2002/01/08 | 17:54:05 | N09E120 | C9.6 | 16.94 | 1794 | 1897 | Y | 14-0.09 |
| 49 | 2002/09/06 | 13:31:49 | S08W120 | ---- | 6.18 | 909 | 937 | HiB | N |
| 50 | 2003/10/21 | 03:54:05 | S16E107 | ---- | 11.67 | 1484 | 1529 | N | 5-1 |
| 51 | 2006/11/06 | 17:54:05 | S05E104 | C8.8 | 14.82 | 1994 | 1997 | N | 4-0.3 |
| **Cycle 24** | | | | | | | | | |
| 1 | 2011/08/09 | 08:12:06 | N17W69 | X6.9 | 12.34 (13.98) | 1610 | 1640 | Y | 14-0.2 |
| 2 | 2011/09/22 | 10:48:06 | N09E89 | X1.4 | 11.47 (13.57) | 1905 | 1905 | Y | 14-0.07 |
| 3 | 2011/10/22 | 10:24:05 | N25W77 | M1.3 | 10.48 (11.49) | 1005 | 1011 | Y | 1.5-0.1 |
| 4 | 2012/01/16 | 03:12:10 | N34E86 | C6.5 | 9.08 (9.08) | 1060 | 1060 | N | 3-0.9 |
| 5 | 2012/01/26 | 04:36:05 | N41W84 | C6.4 | 9.98(10.84) | 1194 | 1195 | HiB | 5-0.5? |
| 6 | 2012/01/27 | 18:27:52 | N27W78 | X1.7 | 9.58 (9.58) | 2508 | 2541 | Y | 14-0.15 |
| 7 | 2012/02/09 | 21:17:36 | N18E80 | B4.2 | 11.21 (11.21) | 659 | 663 | N | N |
| 8 | 2012/02/23 | 08:12:06 | N27W71 | B5.4 | 9.77 (10.26) | 505 | 516 | N | N |
| 9 | 2012/03/04 | 11:00:07 | N19E61 | M2.0 | 8.55 (8.55) | 1306 | 1352 | M | 1-0.2 |
| 10 | 2012/03/13 | 17:36:05 | N17W66 | M7.9 | 13.16 (13.16) | 1884 | 1931 | Y | 14-0.2 |
| 11 | 2012/04/09 | 12:36:07 | N20W65 | C3.9 | 7.75 (8.74) | 921 | 945 | N | 14 - 5 |
| 12 | 2012/05/17 | 01:48:05 | N11W76 | M5.1 | 12.63 (12.63) | 1582 | 1596 | Y | 14-0.3 |
| 13 | 2012/07/19 | 05:24:05 | S13W88 | M7.7 | 16.11 (16.11) | 1631 | 1631 | Y | 5-0.6 |
| 14 | 2012/11/08 | 02:36:06 | N13E89 | M1.7 | 11.81 (11.81) | 855 | 855 | M | Nm |
| 15 | 2012/11/27 | 02:36:05 | N13E68 | C1.3 | 12.38 (12.38) | 844 | 874 | N | N[g] |
| 16 | 2013/05/13 | 02:00:05 | N11E90 | X1.7 | 11.28 (12.55) | 1270 | 1270 | N | 14 - 2 |
| 17 | 2013/05/13 | 16:07:55 | N11E85 | X2.8 | 16.75 (16.75) | 1850 | 1852 | M | 14 -0.3 |
| 18 | 2013/05/14 | 01:25:51 | N08E77 | X3.2 | 19.12 (19.12) | 2625 | 2645 | M | 14-0.24 |
| 19 | 2013/05/15 | 01:48:05 | N12E64 | X1.2 | 10.78 (10.78) | 1366 | 1408 | Y | 14-0.3 |
| 20 | 2013/05/22 | 13:25:50 | N15W70 | M5.0 | 10.78 (12.32) | 1466 | 1491 | Y | 14-0.15 |
| 21 | 2013/09/24 | 20:36:05 | N26E70 | B6.5 | 10.92 (10.92) | 919 | 932 | N | N |
| 22 | 2013/10/25 | 08:12:05 | S08E73 | X1.7 | 7.01 (7.01) | 587 | 599 | N | Nm |
| 23 | 2013/10/25 | 15:12:09 | S06E69 | X2.1 | 10.79 (10.79) | 1081 | 1103 | M | 14-0.2 |
| 24 | 2013/10/28 | 02:24:05 | N04W66 | X1.0 | 9.67 (9.67) | 695 | 726 | M | Nm |
| 25 | 2013/10/29 | 22:00:06 | N05W89 | X2.3 | 11.33 (11.33) | 1001 | 1001 | M | Nm |
| 26 | 2013/11/19 | 10:36:05 | S14W70 | X1.0 | 8.12 (8.12) | 740 | 761 | M | 14-5 |
| 27 | 2014/01/20 | 22:00:05 | S07E67 | C3.6 | 8.58 (8.58) | 721 | 750 | N | 14-8 |
| 28 | 2014/02/20 | 08:00:07 | S15W73 | M3.0 | 6.45 (7.50) | 948 | 960 | Y | 12-7.7 |
| 29 | 2014/02/25 | 01:25:50 | S12E82 | X4.9 | 14.99 (14.99) | 2147 | 2153 | Y | 14-0.1 |
| 30 | 2014/06/10 | 13:30:23 | S17E82 | X1.5 | 11.99 (13.48) | 1469 | 1473 | N | 14-9 |
| 31 | 2014/08/24 | 12:36:05 | S07E75 | M5.9 | 6.65 (6.65) | 551 | 569 | N | Nm |
| 32 | 2015/02/09 | 23:24:05 | N12E61 | M2.4 | 6.16 (7.63) | 1106 | 1148 | N | 14-9? |
| 33 | 2015/03/07 | 22:12:05 | S19E74 | M9.2 | 16.87 (16.87) | 1261 | 1304 | N | 14-8? |
| 34 | 2015/04/23 | 09:36:05 | N12W89 | M1.1 | 9.39 (9.39) | 857 | 864 | M | 3-1 |
| 35 | 2015/05/05 | 22:24:05 | N15E79 | X2.7 | 12.21 (12.21) | 715 | 721 | N | 2-0.5 |
| 36 | 2016/01/01 | 23:24:04 | S25W82 | M2.3 | 9.31 (11.08) | 1730 | 1734 | Y | 1.1–0.3 |
| 37 | 2017/04/18 | 19:48:05 | N14E77 | C5.5 | 14.54 (15.43) | 926 | 932 | N | 2-0.5? |
| 38 | 2017/09/10 | 16:00:05 | S09W90 | X8.2 | 13.87 (14.13) | 3163 | 3163 | Y | 14-0.15 |
| 39 | 2011/10/01 | 20:48:05 | N24E119 | B9.7 | 9.47 (9.47) | 1238 | 1278 | N | NS |
| 40 | 2011/11/04 | 01:25:29 | N20E108 | C5.4 | 8.83 (8.83) | 756 | 772 | HiB | N |
| 41 | 2011/11/17 | 20:36:05 | N18E120 | ----- | 8.22 (9.27) | 1041 | 1202 | N | 12-4 |
| 42 | 2012/01/02 | 15:12:40 | N08W104 | C2.4 | 10.67 (10.67) | 1138 | 1168 | M | N |
| 43 | 2012/01/12 | 08:24:05 | N29E117 | ---- | 12.38 (12.38) | 814 | 914 | N | N |
| 44 | 2012/03/18 | 00:24:05 | N18W116 | B8.1 | 9.36 (9.36) | 1210 | 1346 | N | N |
| 45 | 2012/03/28 | 01:36:07 | N21E116 | B6.4 | 7.40 (8.49) | 1033 | 1149 | N | N |
| 46 | 2012/06/23 | 07:24:05 | N18W101 | C2.7 | 12.99 (12.99) | 1263 | 1315 | N | NS |
| 47 | 2012/08/25 | 16:36:05 | S20E120 | ---- | 4.96 (4.96) | 636 | 734 | N | N |
| 48 | 2012/11/16 | 00:48:06 | S29E120 | ---- | 13.10 (13.10) | 667 | 767 | N | N |
| 49 | 2012/11/16 | 07:24:14 | S14E110 | ---- | 10.38 (11.20) | 775 | 797 | N | N |
| 50 | 2012/11/21 | 04:24:07 | N11W99 | C1.0 | 13.47 (13.47) | 920 | 931 | N | N |
| 51 | 2012/11/23 | 11:00:05 | N16E100 | ---- | 7.11 (7.11) | 342 | 347 | N | N |
| 52 | 2013/04/21 | 07:24:07 | N10W119 | ---- | 6.82 (6.82) | 919 | 1051 | M | 10-5.5 |
| 53 | 2013/05/01 | 03:12:08 | N15E115 | ---- | 7.05 (7.05) | 762 | 841 | N | N |
| 54 | 2013/10/05 | 07:09:51 | S22E118 | ---- | 11.24 (11.24) | 964 | 1092 | HiB | 1-0.25 |
| 55 | 2013/10/11 | 07:24:10 | N21E103 | M1.5 | 14.01 (15.30) | 1200 | 1232 | N | 14-0.2 |
| 56 | 2013/11/07 | 00:00:06 | S11W97 | M1.8 | 11.64 (11.64) | 1033 | 1041 | M | N |
| 57 | 2014/01/06 | 08:00:05 | S15W112 | C2.1 | 12.90 (14.35) | 1402 | 1512 | Y | 14-0.4 |
| 58 | 2014/02/09 | 16:00:06 | S15E103 | M1.0 | 10.20 (11.16) | 908 | 932 | N | 0.8–0.4 |
| 59 | 2014/05/07 | 16:24:05 | S11W100 | M1.2 | 13.68 (13.68) | 923 | 937 | M | 6-0.6? |
| 60 | 2014/05/08 | 03:24:05 | S09W108 | C1.9 | 9.45 (10.34) | 847 | 891 | HiB | 14-3.5 |
| 61 | 2014/09/01 | 22:24:05 | S13E113 | C1.7 | 17.23 (18.68) | 1404 | 1525 | HiB | DGm |
| 62 | 2014/09/26 | 04:28:16 | S13E111 | ----- | 8.16 (9.71) | 1469 | 1565 | N | <1MHz |
| 63 | 2014/10/14 | 18:48:06 | S15E111 | M2.2 | 9.68 (9.68) | 848 | 908 | N | N |

[a]##45-51 in SC 23 and 39-63 in SC 24 are newly added to the list in [30]. [b]Estimated flare size. [c]HiB: high background; M: <10 pfu event. [d]Frequency range of type II bursts; m – metric type II burst.

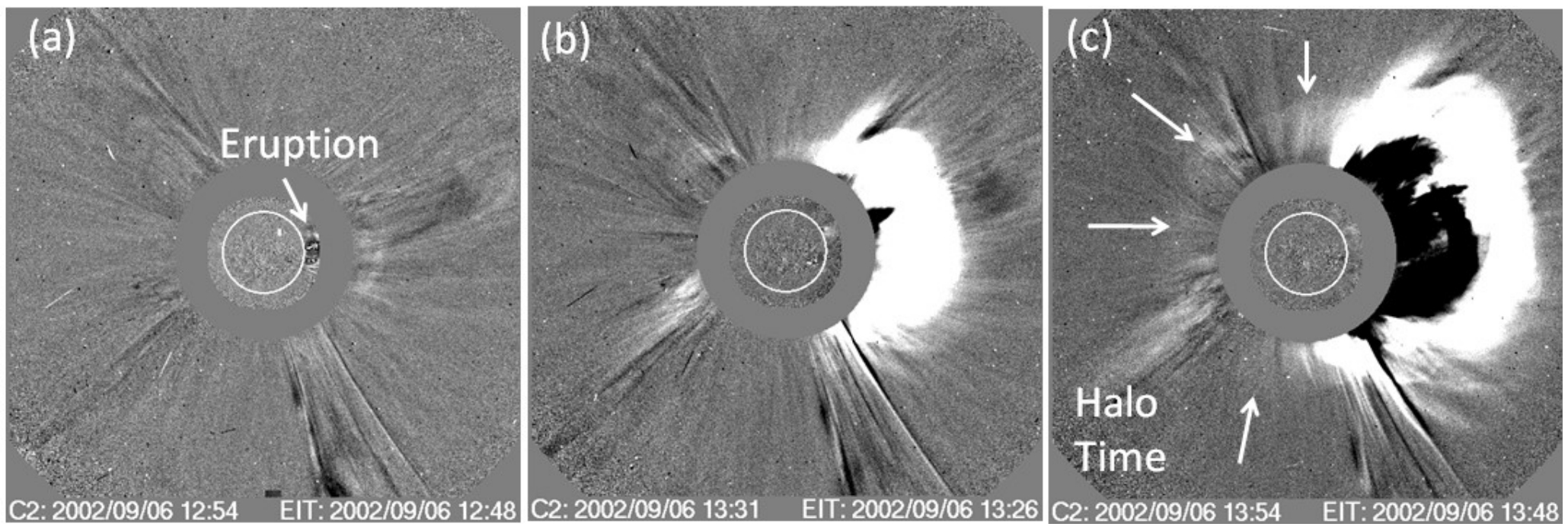

**Figure 8.** A limb halo CME that originated behind the limb. (a) Eruption seen as a EUV disturbance in SOHO/EIT difference image superposed on a LASCO/C2 image. (b) The CME appears in the LASCO/C2 FOV at 13:31 UT and is not a halo yet. (c) Shock can be seen above the east limb and all around the occulting disk (pointed by arrows) at 13:54 UT, which we call the halo time. The CME leading edge (LE) is at a heliocentric distance of 6.18 Rs (halo height). The LE sky-plane ($V_{Sky}$ = 909 km/s) and space speeds ($V_{Sp}$= 937 km/s) are similar (minimal projection effects).

The newly added events in Table 3 are mostly backsided, so the associated flares are mostly of B and C class because of partial occultation. Nine cases in SC 24 and 2 in SC 23 did not have flare size because the flares are fully occulted. The real flare sizes are expected to be larger. The SEP association in the backsided events is poor because these events are poorly connected to an Earth observer. The entries in the SEP column in Table 3 are: Y=Yes, N = No, M = minor (<10 pfu); HiB = high background; NS = no GOES SEP, but STEREO detected it. The association of type II bursts in the backsided events is also poor, but better than SEP association: N = No, Nm = No IP type II but there is metric type II association; if yes, the frequency range is given; DG = data gap; NS = Wind did not detect a type II burst but STEREO detected it.

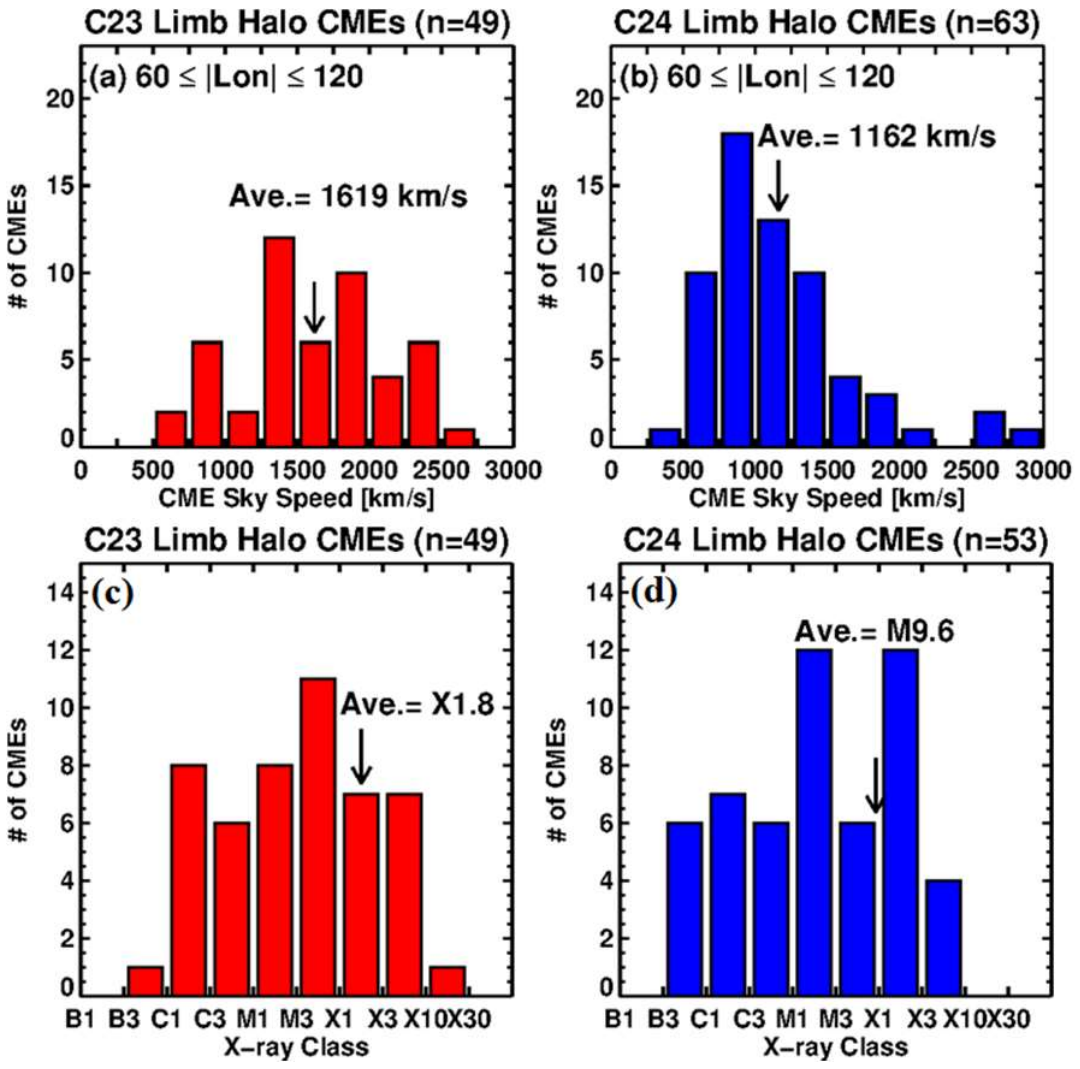

**Figure 9.** CME-speed (a,b) and flare-size (c,d) distributions of the limb halos in SC 23 (red) and 24 (blue) from Table 3. While the speed distributions are different, the flare size distributions are similar. A two-sample KS test gives D = 0.5261 (p = 0.000, Dc = 0.2591) for speeds; D = 0.1405 (p = 0.662, Dc = 0.2695) for flares sizes.

Figure 9 shows the speed and flare-size distributions of limb halos compared between SCs 23 and 24. The average speed of limb halos (1619 ± 147 km/s in SC 23 and 1162 ± 132 km/s in SC 24) are more than a factor of two higher than the corresponding speeds of non-halo limb CMEs (627 ± 33 km/s in SC 23 and 492 ± 32 km/s in SC 24). Limb halos indicate that the CMEs need to be very fast to drive a shock that appears on the opposite limb. The average speed of limb halos in SC-24 dropped significantly by 28%, slightly larger than the drop in the average speed of non-halo limb CMEs (22%). Contrary to the speeds, the flare-size distributions are similar in the two cycles, with average sizes of X1.8 (SC 23) and M9.6 (SC 24) as confirmed by a two-sample KS test. These results are consistent with those obtained using all limb CMEs, except that limb halos are much faster and associated with larger flares.

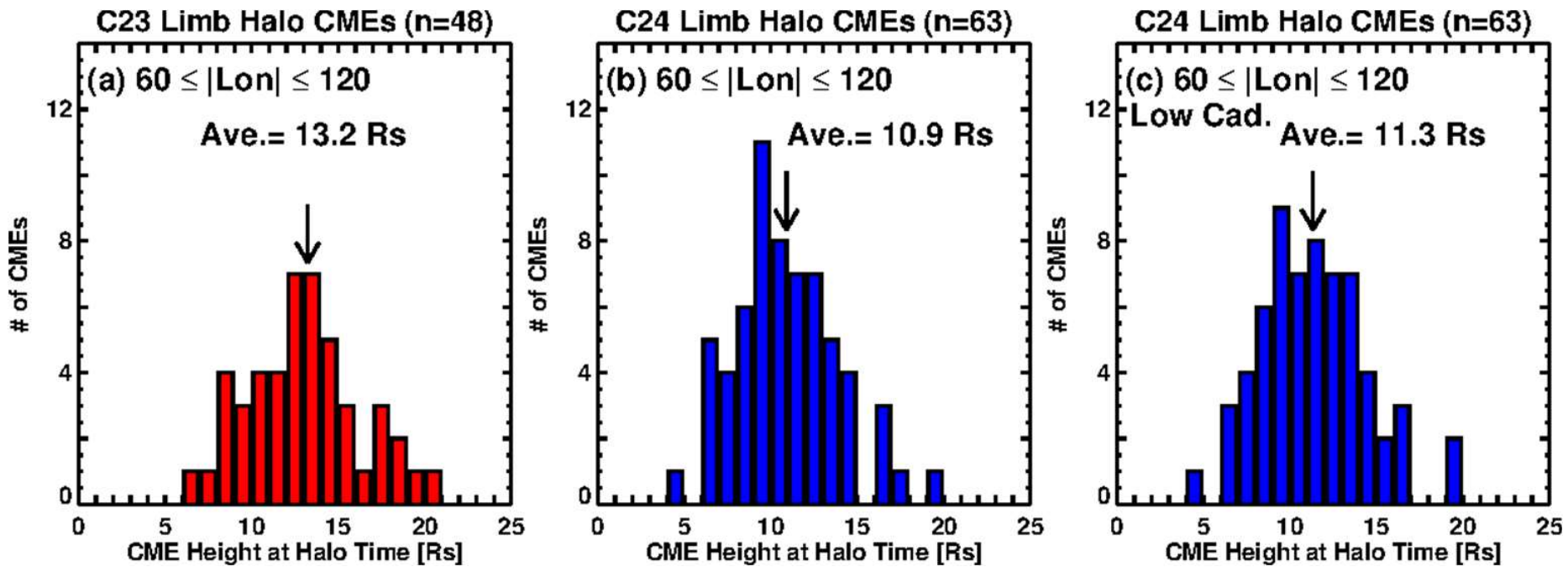

**Figure 10**. Distributions of halo heights in SC 23 (a), in SC 24 with full cadence (b), and in SC 24 with reduced cadence (c). The SC-24 halo heights are significantly smaller than those in SC 23 as indicated by a two-sample KS test that yields D = 0.3462 (p = 0.002, Dc = 0.0367); D >Dc indicates statistical significance at 95% confidence level.

Figure 10 shows the distributions of halo heights in the two cycles. The average halo height in SC 24 (10.9 ± 1.1 Rs) is significantly smaller than that (13.2 ± 1.1 Rs) in SC 23 by 17%. When the SC-24 LASCO cadence is reduced to match that in SC 23, the halo height increases slightly to 11.3 ± 0.8 Rs in SC 24. Taken together with the fact that SC-24 CMEs are slower, this result shows that halos are formed sooner at lower speeds in SC 24. We suggest that this is a direct consequence of the reduced pressure in the heliosphere, which allowed the CMEs in SC 24 expand excessively to become halos.

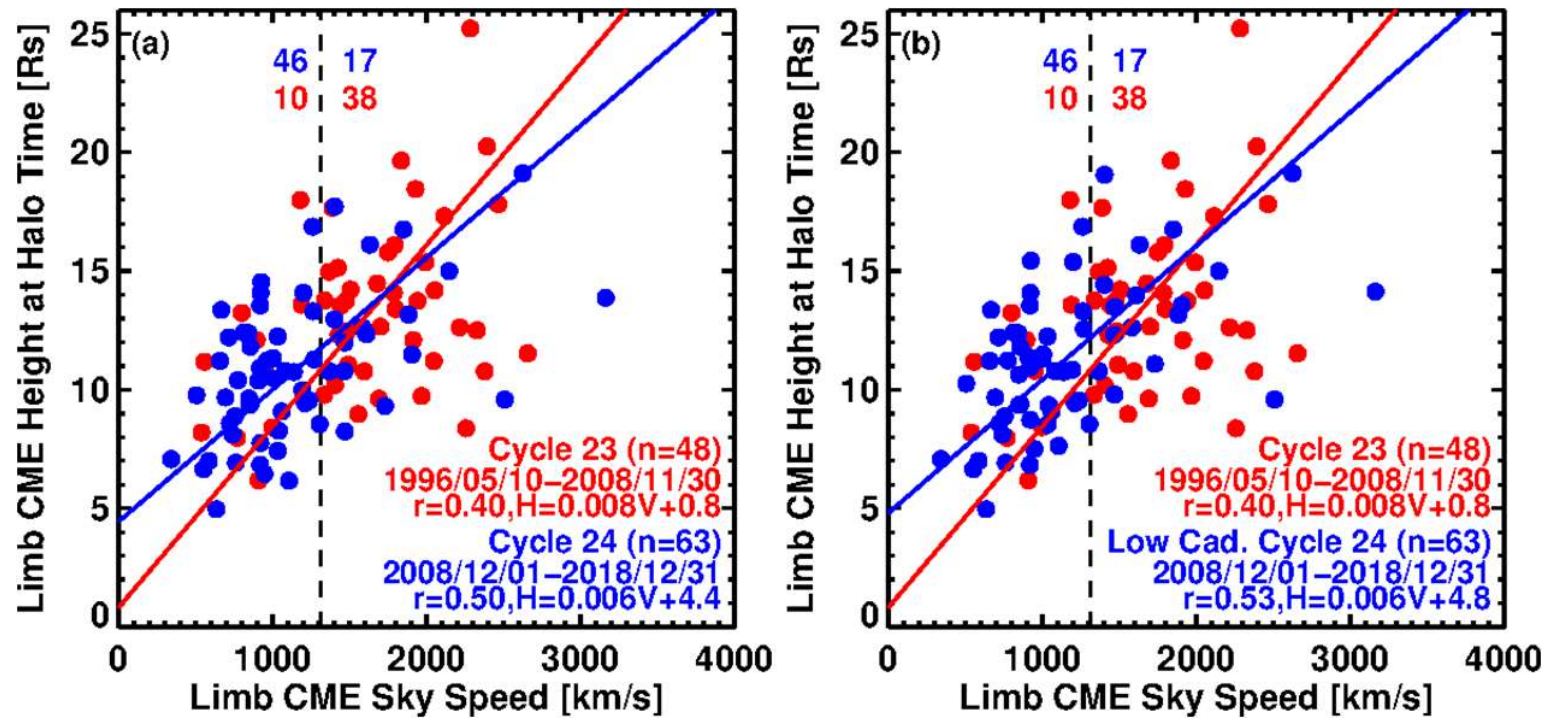

**Figure 11**. Halo height (H) vs. CME speed (V) scatter plots in SC 23 (red) and SC 24 (blue). (a) SC-24 CME data used with actual cadences and (b) with reduced cadence. The correlations are significant. SC-24 (SC 23) CMEs cluster to the left (right) of ~1300 km/s (marked by the vertical dashed lines). The lowest halo height is in the range ~5-6 Rs.

The scatter plots in Fig. 11 show the combined effect of the lower CME speeds and smaller halo heights in SC 24: the blue and red data points are generally at the lower left and upper right, respectively. A speed of ~1300 km/s roughly divides the limb halos into SC-24 and SC-23 events: 46 of the 63 (or 73%) SC-24 halos have speeds <1300 km/s while only 10 of the 48 (or 21%) SC-23 halos have such speeds. The reverse is true for speeds >1300 km/s. The halo heights have a lower cut off in the range 5 – 6 Rs, which is near the outer edge of the LASCO/C2 FOV. This might explain why some catalogs that use only LASCO/C2 images for automatic CME detection may not identify many halo CMEs [16]. The correlation between CME speed and halo height implies faster CMEs becoming halos later for a given cadence.

Finally, there is a high degree of association between limb halos and type II radio bursts (SC 23: 47 out of 51 or 92%, SC 24: 44 out of 63 or 70%). When the 11 STEREO detections in behind-the-limb events are included, the association rate goes up to 55 out of 63 or 87% in SC 24. There was no STEREO data in SC 23, so the type II association rate is a lower limit for that cycle. The association rate is smaller for large SEP events, mainly due to connectivity and high background issues. In SC 23, 19 of the 29 western limb halos (66%) are associated with SEPs, compared to 11 of the 29 (or 42%) in SC 24, roughly following the number of FW CMEs. When minor events are included, the association rate goes up. Many of the behind the limb events had SEP events at one or both of the STEREO spacecraft in SC 24.

## 4. Summary and conclusions

Taking advantage of the availability of CME data over two solar cycles (23 and 24), we compared the occurrence rate of CMEs between the two cycles with reference to the overall solar activity represented by the sunspot number. Higher rate of weak CMEs and lower rate of energetic CMEs occurred in SC 24 than in SC 23. The average speeds of various populations are smaller in SC 24. The level of correlation between CME occurrence rate and SSN is similar in the two cycles, but the slopes are significantly different: there are more CMEs per SSN in SC 24. The only exception is the number of fast and wide CMEs that occur at a slightly lower rate relative to the SSN. The normalized monthly rate of fast and wide CMEs varies very similar to the SSN. The number of ≥C1.0 flares closely follows the SSN indicating that most of such flares originate in sunspot regions. The number of ≥C3.0 flares accompanied by CMEs also varies similar to the SSN, although there is a slight excess in SC 24.

The number of halo CMEs in SC 24 did not drop as much as the SSN did: normalized to the SSN, halos are overabundant in SC 24. This is also true for limb halos. The speed-width relationships in limb CMEs previously compared between partial cycles, continue to be different when compared between full cycles: for a given speed, SC-24 CMEs are wider. A new finding is that SC-24 limb CMEs are slower on average, but still wider. We introduced a new parameter known as halo height – the height at which a CME becomes a halo. The halo hights were determined for limb halos in the two cycles. The average halo height is significantly lower in SC 24. The limb halos are also slower in SC 24 indicating that CMEs become halos sooner at lower speeds in this cycle. This can be understood as a direct consequence of the weak heliospheric state in SC 24 that allows CMEs to expand more and become halos. The soft X-ray flare-size distributions are similar in both limb CMEs and limb halos between the two cycles. We computed the magnetic potential energy at the time of central meridian passage of the source regions that produced the ≥C3.0 flares and found it to be lower in SC 24, consistent with the lower rate of energetic eruptions (fast and wide CMEs). The backreaction of the weak heliosphere seems to be only on CMEs and not on flare sizes. Thus, we can pin down the heliospheric state as the main cause of the CME expansion in SC 24. Furthermore, we made direct comparison between the heliospheric states in SCs 24 and 23 as determined from in-situ solar wind measurements and found most of the solar wind parameters, except the density show significant decline in SC 24.

Space weather events are affected by the reduced solar activity in two ways: (i) the number of fast and wide CMEs decreased, and (ii) the backreaction of the weak heliospheric state in diluting the magnetic content of CMEs (weak geomagnetic storms), and reducing the efficiency of particle acceleration (very few high-energy SEP events). The number of shocks inferred from type II radio bursts faithfully follows the number of fast and wide CMEs. Since the strength of SC 25 is predicted to be similar to that of SC 24, we expect continued milder space weather in SC 25.

**Acknowledgments**
We benefited from the open data policy of SOHO, Wind, STEREO, SDO, and Wind missions. This work was supported by NASA's Living With a Star program. PM was partially supported by NASA grant NNX15AB77G. HX was partially supported by NASA HGI grant NNX17AC47G.